\documentclass[square,aps,preprint,superscriptaddress]{revtex4}
\usepackage{multirow,graphicx}
\begin{document}

\title{Modulation of PEDOT properties via cobalt ferrite nanoparticles: morphology, conjugation length, doping level, structure, and electrical conductivity}

\author{Gabriel Paciaroni}
\affiliation{Universidad de Buenos Aires, Facultad de Ciencias
Exactas y Naturales, Departamento de Qu{\'i}mica Inorg{\'a}nica,
Anal{\'i}tica y Qu{\'i}mica F{\'i}sica, Ciudad Universitaria,
Pabell\'on 2, C1428EHA, Buenos Aires, Argentina}
\affiliation{CONICET- Universidad de Buenos Aires, Instituto de
Qu{\'i}mica F{\'i}sica de Materiales, Ambiente y Energ{\'i}a
(INQUIMAE), Ciudad Universitaria, Pabell\'on 2, C1428EHA, Buenos
Aires, Argentina} \affiliation{Universidad de Buenos Aires, Facultad
de Ciencias Exactas y Naturales, Departamento de F{\'i}sica,
Laboratorio de Bajas Temperaturas, Ciudad Universitaria, Pabell\'on
1, C1428EHA, Buenos Aires, Argentina} \affiliation{CONICET -
Universidad de Buenos Aires, Instituto de F{\'i}sica de Buenos Aires
(IFIBA), Ciudad Universitaria, Pabell\'on 1, C1428EHA, Buenos Aires,
Argentina}

\author{Mar\'{i}a Ana Castro}
\affiliation{Universidad de Buenos Aires, Facultad de Ciencias
Exactas y Naturales, Departamento de Qu{\'i}mica Inorg{\'a}nica,
Anal{\'i}tica y Qu{\'i}mica F{\'i}sica, Ciudad Universitaria,
Pabell\'on 2, C1428EHA, Buenos Aires, Argentina}
\affiliation{CONICET- Universidad de Buenos Aires, Instituto de
Qu{\'i}mica F{\'i}sica de Materiales, Ambiente y Energ{\'i}a
(INQUIMAE), Ciudad Universitaria, Pabell\'on 2, C1428EHA, Buenos
Aires, Argentina}

\author{Carlos Acha}
\affiliation{Universidad de Buenos Aires, Facultad de Ciencias
Exactas y Naturales, Departamento de F{\'i}sica, Laboratorio de
Bajas Temperaturas, Ciudad Universitaria, Pabell\'on 1, C1428EHA,
Buenos Aires, Argentina} \affiliation{CONICET - Universidad de
Buenos Aires, Instituto de F{\'i}sica de Buenos Aires (IFIBA),
Ciudad Universitaria, Pabell\'on 1, C1428EHA, Buenos Aires,
Argentina}

\author{Paula Soledad Antonel}
\thanks{corresponding author (sole@qi.fcen.uba.ar)}
\affiliation{Universidad de Buenos Aires, Facultad de Ciencias
Exactas y Naturales, Departamento de Qu{\'i}mica Inorg{\'a}nica,
Anal{\'i}tica y Qu{\'i}mica F{\'i}sica, Ciudad Universitaria,
Pabell\'on 2, C1428EHA, Buenos Aires, Argentina}
\affiliation{CONICET- Universidad de Buenos Aires, Instituto de
Qu{\'i}mica F{\'i}sica de Materiales, Ambiente y Energ{\'i}a
(INQUIMAE), Ciudad Universitaria, Pabell\'on 2, C1428EHA, Buenos
Aires, Argentina}

\begin{abstract}
Composite materials based on Poly(3,4-ethylenedioxythiophene)
(PEDOT) and CoFe2O4 magnetic nanoparticles (NP) were synthesized by
chemical oxidative polymerization with varying monomer and
surfactant (DBSA) concentrations, and were compared to PEDOT samples
synthesized without NP. Electrical conductivity measurements were
performed, which revealed that the composites are more conductive
than the pure PEDOT samples, with this effect depending on EDOT and
DBSA contents. Characterizations by SEM and TEM microscopies,
UV-Vis, FTIR and Raman spectroscopies, X-ray diffraction and dynamic
light scattering were carried out in order to associate the
morphology and structure of these materials to their electrical
conductivity, and to explain how EDOT and DBSA concentrations, and
also the presence of NP, affects those properties. It was found that
the NP play a significant role in the polymerization of EDOT,
influencing the formation and arrangement of polymer chains, as well
as their conjugation length, oxidation state, and resonant
structures. These effects are also dependent on the DBSA content. To
describe the conductivity of the composites, a two-phase model based
on general effective media theory was introduced. The analysis
revealed that, at low reactant concentrations, the NP increase the
conductivity of the adjacent PEDOT by over two orders of magnitude.

\end{abstract}

\maketitle

\section{Introduction}
\label{} Conducting polymers possess interesting chemical,
electrical, mechanical and physical properties, which make them
attractive materials in a number of applications such as in organic
and flexible electronics, energy storage systems, electromagnetic
shielding and biocompatible devices \cite{Kumar1998, Balint2014,
Chandrasekhar2018, Ouyang2021}. In particular,
Poly(3,4-ethylenedioxythiophene) (PEDOT) is a conducting polymer
with excellent environmental stability, high electrical conductivity
and high transparency in the visible range \cite{Aleshin1999,
Pei1994, Chiu2005}. The synthesis of PEDOT can be achieved through
several methods, including electrochemical polymerization and
chemical oxidative polymerization \cite{Jiang2012, Paradee2013,
Kirchmeyer2005, Gueye2020, Tolstopyatova2009, Im2007, Han2006,
Zhao2014}, being the later particularly advantageous for large-scale
production of nanostructured PEDOT and for device fabrication. Among
chemical synthesis methods of PEDOT, the dispersion or emulsion
polymerization is one of the most employed, because it overcomes the
problem of the low solubility of EDOT in water \cite{Tumov2023,
Choi2004, Chen2008, Lei2005}. Furthermore, the emulsion
polymerization carried out with, for example, different surfactant
concentrations gives the possibility of tuning and customizing the
electrical conductivity of PEDOT and other physical and chemical
properties, such as the morphology, structure, oxidation state and
doping level \cite{Choi2004, Lei2005, Oh2002}. A emulsifier
frequently used in the synthesis of PEDOT is dodecylbenzenesulfonic
acid (DBSA), which also acts as dopant \cite{Chen2008, Chutia2014,
Tumov2023}. Besides, PEDOT doped with DBSA appears in literature as
an improved substitution of PEDOT:PSS and offers several benefits in
terms of biocompatibility and improved film processability, making
it a promising material for applications in bioelectronics
\cite{Tumov2023}. As DBSA acts as PEDOT dopant, the morphology and
the electrical conductivity of the resulting material can be tuned
by controlling DBSA concentration and DBSA:EDOT molar ratio
\cite{Chutia2014}. Although several reports exist on the emulsion
polymerization of EDOT using DBSA as a surfactant, a systematic
study on the effect of DBSA concentration on the morphology and
structure of the resulting polymer, and their impact on the
electrical conductivity of PEDOT, has not been conducted yet.

On the other hand, hybrid materials, or composites, based on
inorganic nanoparticles (NP) and organic polymers are of special
interest, due to the possibility of combining and improving the
properties of the respective precursor materials \cite{Iqbal2018,
Jeon2010}. Particularly, the combination of magnetic NP and
conducting polymers results in composites which are sensitive to
both electric and magnetic fields, with proposed applications in
spintronics, sensors and memories, between others \cite{Khan2024,
MuozBonilla2016, K2023}. A usual process for obtaining these
composites consists in carrying out the polymerization in the
presence of a dispersion of NP \cite{De2009, Ohlan2010,
LanusMendezElizalde2020, Elizalde2022, Landa2021}. At this point it
is important to note that the presence of those NP, along with the
other reactants used for the polymerization, may affect the growth
of the polymer. Under those conditions, the resulting composite
would differ from a material created by merely mixing the NP with
the polymer post-synthesis.

In order to provide tools for the aforementioned applications it is
of great importance to study the influence of both the conducting
polymer and the NP on the properties of the resulting composite. In
our previous works we investigated the correlation between the
composition and the electrical and magnetic properties of composite
materials based on magnetic NP and different conducting polymers
\cite{LanusMendezElizalde2020, Elizalde2022, Landa2021, Antonel2015,
GarcaSaggin2020, Resta2013, Resta2017}. We found that the obtained
properties differ from the ones expected from a simple dilution and
that it is possible to control the morphology, electrical
conductivity, and magnetic behavior of those nanostructured
composites by the NP:monomer molar ratio employed in their
synthesis. Moreover, the non-trivial magnetic behavior of our
composites was rationalized in terms of the change of the type and
magnitude of the magnetic interactions between NP
\cite{LanusMendezElizalde2020, Landa2021, Elizalde2022,
Antonel2015}, revealing the importance of NP-NP distances on the
final properties. As just outlined, these studies were focused on
the effect that the conducting polymer has on NP properties, mainly
in their magnetic behavior. On the other hand, and in order to have
a better understanding of the composite properties, it is also
necessary to investigate if the presence of NP affects the
conducting polymer, particularly its structure (doping, oxidation
state), morphology and electrical conductivity, when compared to
pure polymer.

To the best of our knowledge, there are few reports that
characterize PEDOT and magnetic NP composites by studying the
influence of both NP and other reactants in the electrical
conductivity of the resulting material. The way in which the polymer
and the NP interact must also be investigated through both
morphology and structure studies of the composites, for which
microscopic and spectroscopic characterizations, respectively, are
needed. Those techniques are not usually employed in a complementary
fashion in the context of these materials. Additionally, it is not
common to combine different types of spectroscopies (UV-vis, FTIR,
Raman) in order to gain deeper insight about the chemical structure
of this polymer, and relate it to its electrical conductivity.
Furthermore, it is necessary to have a better understanding of the
electrical transport in these composites, and physical models based
on percolation and general effective media theories appear as
attractive tools \cite{McLachlan1990, Stcker2012, CruzEstrada2002}.

In this work, composites based on PEDOT and magnetic cobalt ferrite
(CoFe$_2$O$_4$) NP were chemically synthesized, along with pure
PEDOT samples, employing different concentrations of the reactants
EDOT and DBSA. Electron microscopic and spectroscopic (with emphasis
on Raman spectroscopy) characterizations were carried out in order
to study their morphology and structure, with the objective of
explaining the influence of the presence of NP and DBSA in their
macroscopic properties, particularly in their electrical
conductivity. Physical mixtures of PEDOT and CoFe$_2$O$_4$ NP in
different proportions were also prepared for comparative studies, in
order to gain information about the interactions that occur between
the particles and the polymer in the composites.

\section{Experimental}
AR grade chemicals, supplied by Sigma-Aldrich, and high purity water
(18 M$\Omega$ cm) from a Milli-Q system were employed throughout.
Ethylenedioxythiophene (EDOT) was used as received.

\subsection{Synthesis of cobalt ferrite nanoparticles}
The synthesis of  CoFe$_2$O$_4$ NP was performed by the
co-precipitation method \cite{Antonel2015}. Briefly, 22.25 mL of a
solution containing 0.450 M  FeCl$_3\cdot_6$H$_2$O and 0.225 M
CoCl$_2\cdot_6$H$_2$O (2:1 Fe(III)-Co(II) molar ratio), in 0.4 M
HCl, was added dropwise to 200 mL of 1.5 M NaOH under high speed
mechanical stirring. The synthesis temperature was set at 80
$^{\circ}$C, using a water-jacketed reaction vessel with a
circulating thermostatic bath. Dark brown  CoFe$_2$O$_4$
nanoparticles precipitated immediately after the first drops of the
Fe(III)-Co(II) solution. The temperature of synthesis and the
high-speed mechanical stirring were kept constant during the
addition of the cationic solution. After the addition of this
solution, the reaction media was maintained at 80º C, at high-speed
stirring, for 2 hrs. The CoFe$_2$O$_4$ nanoparticles were separated
by centrifugation at 8000 G during 20 minutes at room temperature.
The pellet was washed with Milli-Q water, repeating the cycles of
washing-centrifugation until neutral pH of the supernatant was
reached. Finally, the CoFe$_2$O$_4$ nanoparticles were dried using a
vacuum oven at 40$^{\circ}$ C during 24 hrs.

\subsection{Synthesis of PEDOT: CoFe$_2$O$_4$  composites and PEDOT:DBSA}
PEDOT: CoFe$_2$O$_4$  composites were synthesized following Lanus
Mendez Elizalde et al \cite{LanusMendezElizalde2020} by varying the
reactants concentrations in order to study their influence on the
final properties. First,  CoFe$_2$O$_4$  NP were added to a
dodecylbenzenesulfonic acid (DBSA) solution of a certain
concentration (11, 33 or 100 mM) to form an aqueous micellar
dispersion. This was carried out by ultrasound treatment and high
speed mechanical stirring for 30 minutes and, after that, a brown
emulsion was obtained. Then, ethylenedioxythiophene (EDOT) monomer
was added in a 1:1 EDOT:DBSA molar ratio for 33 and 100 mM DBSA, and
3:1 for 11 mM DBSA. In all cases, the EDOT:NP molar ratio was fixed
in 2, keeping the reaction mixture for 1 hr in the same conditions
of high speed mechanical stirring and ultrasound treatment. After
that, a required quantity of ammonium persulfate (APS) in a 1:1
EDOT:APS molar ratio was added to the reaction mixture and the EDOT
emulsion polymerization reaction was allowed to proceed for 3 hrs
under the same conditions described before, maintaining room
temperature. Finally, the resulting products were demulsified with
an equal volume of isopropyl alcohol. The dark blue colored
composite samples were separated by centrifugation at 8000 G during
5 minutes at room temperature and washed with isopropyl alcohol.
Then, alternating cycles of washing-centrifugation in ethanol and
MilliQ water were repeated in order to remove excess oligomers and
DBSA. Finally, the samples were dried under vacuum at room
temperature for 24 hrs. As it was mentioned above, the synthesis was
performed for different reactants concentrations, and composite
samples are designated as $C_{dil}$ (11 mM DBSA, 33 mM EDOT),
$C_{mid}$ (33 mM DBSA, 33 mM EDOT) and $C_{conc}$ (100 mM DBSA, 100
mM EDOT). In addition, PEDOT:DBSA samples without  CoFe$_2$O$_4$  NP
were synthesized applying the same procedure described above for
comparative studies. Those samples are accordingly named as
$P_{dil}$, $P_{mid}$ and $P_{conc}$.

\subsection{Preparation of  CoFe$_2$O$_4$ -PEDOT mixtures}
For a comparison study, samples of mixtures of  CoFe$_2$O$_4$  NP
and PEDOT:DBSA of various wt\% were also prepared. To this end,
certain amounts of PEDOT:DBSA synthesized with concentrated
conditions, $P_{conc}$, and  CoFe$_2$O$_4$  NP were intimately mixed
using a mortar in the following NP wt\%: 13, 36, 72 and 90. The
mixtures were designated as $M_{13}$, $M_{36}$, $M_{72}$ and
$M_{90}$, respectively.

\subsection{X-Ray diffraction (XRD) measurements}
X-Ray powder diffraction analysis was performed with a Panalytical
Empyrean diffractometer, using Cu K$\alpha$ radiation ($\lambda$ =
0.154056 nm). The Scherrer equation (\ref{eq:rx}) was used to
determine the average crystallite size $d_c$:
\begin{equation}
    d_c = \frac{A\lambda}{\beta cos(\theta)},
    \label{eq:rx}
\end{equation}
where $A$ is a shape factor (taken as 0.9), $\beta$ is the full
width at half maximum of the peak, and $\theta$ is the corresponding
Bragg angle. The results of the XRD measurements are shown in Fig.
S1 (Supplementary Information).

\subsection{Dynamic light scattering (DLS)}
Dynamic light scattering (DLS) measurements were performed at room
temperature (25 $^{\circ}$C) using a Horiba SZ-100 analyzer equipped
with a 532 nm laser, operating at a scattering angle of 173
$^{\circ}$. This technique was employed to evaluate the stability of
$\mathrm{CoFe_2O_4}$ NP dispersions in aqueous DBSA solution. In
order to replicate the synthesis conditions of the composites, a
defined mass of NP was dispersed in a 33 mM DBSA aqueous solution,
corresponding to a molar ratio of $\mathrm{DBSA:CoFe_2O_4}$ of 2:1.
This dispersion was mechanically stirred and subjected to ultrasonic
treatment for 30 minutes to ensure proper dispersion of the NP. For
comparison, a second dispersion containing the same amount of NP in
an equivalent volume of ultrapure water (without DBSA) was prepared,
in order to evaluate the effect of DBSA dispersion stability. Prior
to analysis, both dispersions were filtered through a 0.2
$\mathrm{\mu m}$ pore-size membrane. Each sample was measured three
consecutive times, with 10-minute intervals between measurements, in
order to assess the temporal stability of the NP hydrodynamic
diameter.

The DLS results are presented in Figure S2 (Supplementary
Information).

\subsection{Electron microscopy studies}
The particle size and morphology of NP, PEDOT:DBSA and composites
were studied by Transmission Electron Microscopy (TEM) and Scanning
Electron Microscopy (SEM). TEM observation was carried out using a
Zeiss EM 109T microscope equipped with a Gatan ES1000W digital
camera. SEM analysis was performed using a Zeiss Crossbeam 340
microscope. The specimens for both techniques were prepared by
dispersing the solids in ethanol. For TEM, one drop of the ethanol
suspension was placed onto a TEM grid, whereas for SEM 10 $\mu$L of
the same suspension were dropped on one side of a silicon wafer. In
both cases, the specimens were allowed to dry at room temperature
for 24 hrs.

\subsection{Spectroscopic characterizations}
For the absorbance measurements in the UV-vis region, PEDOT:DBSA and
composites powders were dissolved in N-methylpyrrolidinone (NMP) and
the UV-vis spectra were recorded in the range 280-850 nm using a
Shimadzu PC3101 spectrophotometer, under computer control.

For the Fourier Transform Infrared Spectroscopy (FTIR)
characterization, the IR spectra of PEDOT:DBSA, NP and composites
were recorded using a FTIR Nicolet 8700 spectrometer, in the range
400-4000 cm$^{-1}$, with a resolution of 4 cm$^{-1}$. The samples
were pressed into pellets prepared dispersing 0.5 mg of each one in
150 mg of KBr. For each sample, 32 scans were accumulated.

Finally, for Raman characterization, the Raman spectra of the
samples were collected using a Horiba LabRam HR Evolution dispersive
spectrometer and microscope equipped with a 633 nm and a 532 nm CVi
Melles Griot laser operating at 0.5 - 2.0 mW, and a CCD Synapse
detector cooled thermoelectrically with a Peltier device. Raman
shifts were measured with a resolution of 0.3 cm$^{-1}$ and, for
each sample, 3 scans were accumulated.

\subsection{Electrical conductivity measurements}
In order to measure the electrical conductivity of the polymer,
composites and mixtures samples, the powders were pressed into
rectangular pellets of dimensions (5 $\times$ 1 $\times$ 0.5)
mm$^3$. Pelletization was carried out with a hydraulic press at 5
kg/cm$^2$ and room temperature (20 $^{\circ}$C). Four contacts were
attached on the surface of each pellet with silver paint, and
four-terminal resistance measurements were performed with a Lock-in
Amplifier, applying low current densities to avoid thermal effects.
Resistivity was calculated from the geometric dimensions of the
pellets.

\section{Results and discussion}

\subsection{Morphology}

\begin{figure}[h]
    \centering
    \includegraphics[width=\textwidth]{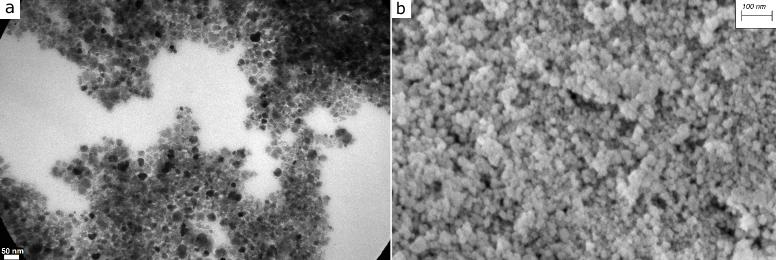}
    \caption{(a) TEM and (b) SEM images of the  CoFe$_2$O$_4$  NP.}
    \label{fig:NPs}
\end{figure}

In Fig. \ref{fig:NPs}, TEM and SEM images of  CoFe$_2$O$_4$  NP are
shown. The NP show a spherical-like shape, similar to those
synthesized in Antonel et al \cite{Antonel2015} with the same
procedure. Using the TEM image, the mean diameter of the
nanoparticles was found to be (15.1 $\pm$ 3.4) nm, measured on the
ImageJ software on 200 particles.

\begin{figure}[h]
    \centering
    \includegraphics[width=\textwidth]{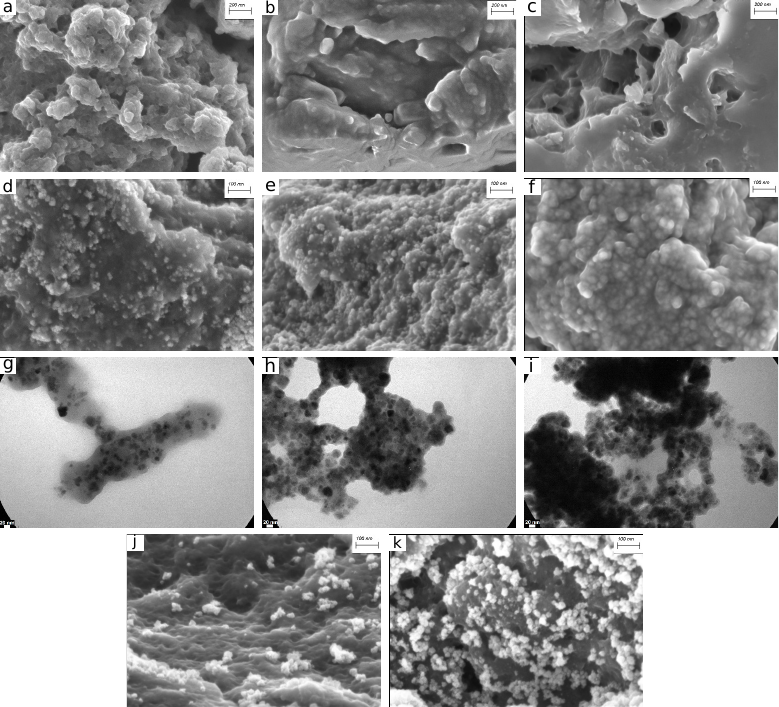}
    \caption{(a), (b), (c): SEM images of the polymer samples $P_{conc}$, $P_{mid}$ and $P_{dil}$. (d), (e), (f): SEM images of the composites $C_{conc}$, $C_{mid}$ and $C_{dil}$. (g), (h), (i): TEM images of the composites $C_{conc}$, $C_{mid}$ and $C_{dil}$. (j), (k): SEM images of the mixtures $M_{36}$ and $M_{90}$.}
    \label{fig:Micros}
\end{figure}

SEM images of the samples containing PEDOT without
 CoFe$_2$O$_4$  NP are shown in Fig. \ref{fig:Micros}a-c. A
distinct morphology can be identified in each of the PEDOT samples,
which could be attributed to the different DBSA molar concentrations
employed in those synthesis. It is known that the critical micelle
concentration, CMC, for DBSA is 29 mM \cite{Paradee2013}, thus no
micelles formation is expected below this concentration value.
Moreover, increasing DBSA molar concentration above its CMC results
in the formation of more and larger micelles \cite{Petrenko2010,
Paradee2013}. The sample with the highest reactant concentrations,
$P_{conc}$, where [DBSA] = 100 mM, that is, much higher than the
CMC, exhibits a globular-like morphology, where the polymer is
aggregated in clusters. This morphology is the result of a
polymerization that occurs within the micelle core, with the later
agglomeration of PEDOT particles due to the micelle fusion. On the
other hand, the sample with medium reactant concentrations,
$P_{mid}$, shows smaller and more dispersed globular structures,
with most of the regions of the polymer appearing smoother. In this
case, DBSA molar concentration (33 mM) is only slightly higher than
the CMC and, therefore, the observed morphology is consistent with a
low amount of micelles in the reaction medium which does not favor
the agglomeration of PEDOT particles. Finally, the sample with the
lowest concentration of DBSA, $P_{dil}$, where [DBSA] = 11 mM, that
is, lower than the CMC, has a flake-like and more irregular
morphology which is the result of no micelle formation in the
reaction medium \cite{Chen2008}.

In Fig. \ref{fig:Micros}d-f, the SEM images of the composites
$C_{conc}$, $C_{mid}$ and $C_{dil}$ are shown. In all cases,
 CoFe$_2$O$_4$  NP are seen embedded in the PEDOT matrix and
distributed uniformly, reflecting the high stability of the NP
dispersions in the DBSA aqueous solution used during composite
synthesis. Indeed, DLS measurements (Fig. S2-Supplementary
Information) show that $\mathrm{CoFe_2O_4}$ NP form a highly stable
and homogeneous dispersion, with no observable increase in
agglomeration over time, within experimental error. These results
confirm the crucial role of DBSA in stabilizing the NP dispersion
and ensuring homogeneous distribution during the synthesis of the
composites. SEM images also reveal a decrease in the
particle-particle separation as both EDOT and DBSA molar
concentrations also decrease, consistent with a lower proportion of
PEDOT, as discussed below. Besides, the three composites show a more
compacted morphology than the corresponding pure PEDOT samples, with
less empty space and holes between the polymer chains. This
observation, together with the fact that the polymer mostly appears
covering  CoFe$_2$O$_4$  NPs, suggests that the presence of those
NPs in the reaction media directs the polymerization of EDOT,
promoting the formation of PEDOT around them. Moreover, in all cases
the polymerization yield increases significantly when
 CoFe$_2$O$_4$  NP are included in the reaction medium.
Therefore, a surface catalytic effect of  CoFe$_2$O$_4$  NP on the
polymerization of EDOT is evidenced, a result that is in agreement
with our previous works of PEDOT: CoFe$_2$O$_4$  and PEDOT:
Fe$_3$O$_4$ composites \cite{LanusMendezElizalde2020, Elizalde2022}.
It is also worth mentioning that, by comparing each synthesis
condition (concentrated, medium, diluted) in the absence and in the
presence of  CoFe$_2$O$_4$  NP, the largest difference in the degree
of polymerization was observed for the diluted conditions. In other
words, considerably much more material was obtained for $C_{dil}$ in
comparison with $P_{dil}$, despite having the same EDOT and DBSA
concentrations. This supports the idea that the nanoparticles act as
a catalyst for the polymerization reaction, providing additional
surface area for PEDOT to form on. This effect is magnified in the
diluted synthesis probably because the scarcity of DBSA causes most
of the polymerization to take place near the nanoparticles.
Conversely, at higher DBSA concentrations, polymerization can
proceed independently of the NP. At this point, it is worth noting
that although specific DBSA nanoparticle interactions might indeed
contribute to the observed behavior, our results do not indicate any
suppression of the catalytic activity of $\mathrm{CoFe_2O_4}$
nanoparticles for EDOT polymerization across the entire range of
DBSA concentrations explored in this work.

Despite the fact that the three syntheses were performed by using
the same EDOT: CoFe$_2$O$_4$  molar ratio, the changes in the
particle-particle separations indicate that a higher proportion of
PEDOT is obtained when both EDOT and DBSA molar concentrations
increase, as in our previous work \cite{Elizalde2022}. Specifically,
$C_{conc}$ sample consists of a composite with the highest
proportion of PEDOT, where some regions of the polymer without
 CoFe$_2$O$_4$  NP are observed. On the other hand, in $C_{dil}$
the  CoFe$_2$O$_4$  NP appear closely aggregated, and covered by
PEDOT, whereas $C_{mid}$ shows an intermediate morphology, more
similar to $C_{conc}$. In order to understand these changes in the
composite morphology, specifically the increase of the proportion of
PEDOT as both EDOT and DBSA molar concentrations in the reaction
medium increase, the influence of those reactants on the chemical
oxidative polymerization of EDOT should be considered. First, by
comparing the synthesis conditions of $C_{dil}$ and $C_{mid}$, DBSA
molar concentration is three times lower in the former, whereas EDOT
molar concentration is the same. It is worth mentioning that, as
DBSA is used as surfactant, the degree of polymerization of EDOT
increases with the concentration of the emulsifier, specifically for
values up to 0.1 M in the case of DBSA, due to a higher amount of
micelles in the reaction medium \cite{Choi2004}. In other words, for
increasing DBSA molar concentrations an increased polymerization
rate is observed \cite{Chen2008}, which results in a higher PEDOT
proportion in $C_{mid}$ in comparison with $C_{dil}$. Then, by
comparing the synthesis conditions of $C_{conc}$ and $C_{mid}$, even
though EDOT:DBSA molar ratio is the same in both synthesis, EDOT and
DBSA molar concentrations are three times higher in the former. On
the one hand, a higher [DBSA] results in an increased degree of EDOT
polymerization, as it was discussed above. On the other hand, it is
known that increasing EDOT molar concentration also enhances the
polymerization rate, giving rise to a higher amount of formed PEDOT
\cite{Han2015, Ha2004}. Therefore, in $C_{conc}$ the highest
proportion of PEDOT is obtained, with the  CoFe$_2$O$_4$  NP more
separated in comparison with $C_{mid}$ and $C_{dil}$.

In Fig. \ref{fig:Micros}d-f also globular structures, mainly in the
regions with  CoFe$_2$O$_4$  NP, are observed, and their mean
diameters were estimated from those SEM images. Unlike in the TEM
images, where the diameter of the nanoparticles can be more clearly
distinguished, the agglomerates in the SEM images may correspond to
NP covered by a polymer shell. These mean diameters were estimated
in (21 $\pm$ 5) nm, (20 $\pm$ 4) nm and (26 $\pm$ 6) nm for the
composites $C_{conc}$, $C_{mid}$ and $C_{dil}$, respectively, which
are 50 - 70 $\%$ larger than the diameter of the nanoparticles
measured in TEM images. This suggests that what is seen as
“apparent NP” in the SEM images of the composites are, in fact,
 CoFe$_2$O$_4$  NP surrounded by a PEDOT layer. This layer is
thicker in $C_{dil}$ because the polymer is preferably formed closer
to the NP, thus the polymerization is more guided by them in the
diluted conditions where [DBSA] is really low.

The NP acting as seed for EDOT polymerization is also evidenced in
the TEM images of the composites shown in Fig. \ref{fig:Micros}g-i.
For the three samples most of the PEDOT and the  CoFe$_2$O$_4$  NP
are closely associated, that is, there are no significant regions
containing PEDOT without NP or vice versa. The fact that the polymer
can be seen covering the NP further confirms that the polymer
preferably grows around them in the three cases, as it was deduced
above. Besides, the decrease in the PEDOT proportion as the
concentrations in the reaction medium are reduced can also be
appreciated.

Using the TEM images, the mean diameters of the  CoFe$_2$O$_4$  NP
in the composites were measured with the aid of ImageJ software.
They resulted in (14.3 $\pm$ 2.8) nm for $C_{conc}$, (14.8 $\pm$
2.3) nm for $C_{mid}$ and (15.4 $\pm$ 2.1) nm for $C_{dil}$. This
agreement with the diameter of the pure  CoFe$_2$O$_4$  NP indicates
that the size of the nanoparticles is not affected by the
polymerization reaction and its acidic conditions, and confirms the
role of DBSA as an effective NP protector
\cite{LanusMendezElizalde2020, Elizalde2022}.

For comparative studies, in Fig. \ref{fig:Micros}j-k SEM images of
two prepared PEDOT+NP mixtures are shown. They are compared to
$P_{conc}$ (Fig. \ref{fig:Micros}a) and to $C_{conc}$ (Fig.
\ref{fig:Micros}d) since the polymer used in the mixtures was
prepared under concentrated synthesis conditions. As it can be seen,
there are marked differences in the morphology since it is evident
that the NP in the mixtures are disaggregated from the polymer
matrix, whereas in $C_{conc}$ they are embedded in it. Besides, in
the mixtures,  CoFe$_2$O$_4$  NP appear agglomerated forming non
uniform clusters, due to the magnetic dipolar interactions between
them. Moreover, the comparison of the morphologies of $M_{36}$ and
$M_{90}$ with that of $C_{conc}$ confirms that PEDOT grows around
 CoFe$_2$O$_4$  NP, that is, those particles direct the
polymerization reaction.

All these differences in the morphologies will be resumed later in
this manuscript, mainly in the context of the electrical
conductivity of the studied samples.

\subsection{UV-Vis spectroscopy}
\begin{figure}[h]
    \centering
    \includegraphics[width=10cm]{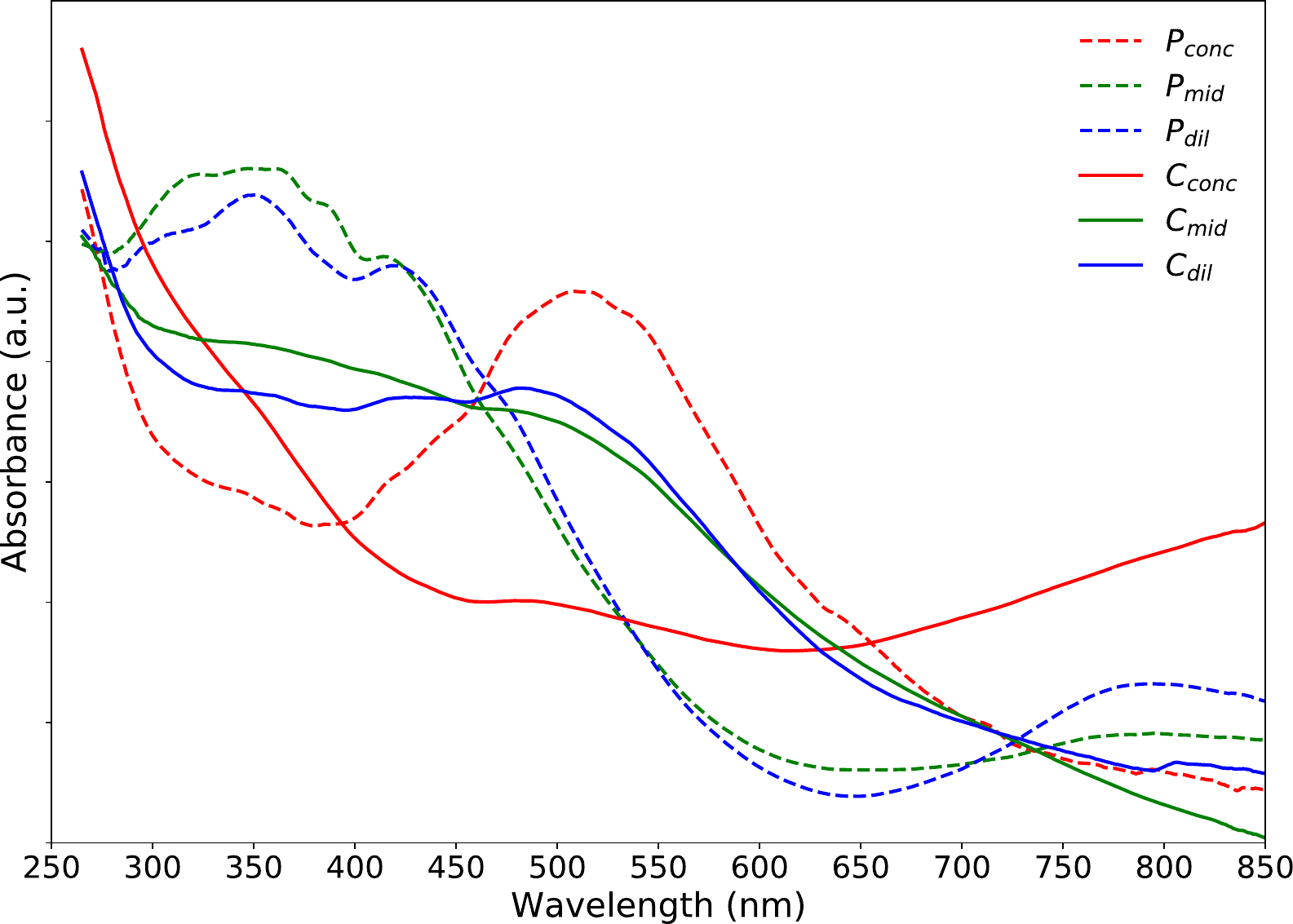}
    \caption{UV-vis spectra of N-methylpyrrolidinone (NMP) solutions of pure PEDOT samples and PEDOT: CoFe$_2$O$_4$  composites.}
    \label{fig:UV}
\end{figure}

The UV-Vis spectra for the samples $P_{conc}$, $P_{mid}$, $P_{dil}$,
$C_{conc}$, $C_{mid}$ and $C_{dil}$ are shown in Fig. \ref{fig:UV}.
First, UV-Vis absorption spectra of $P_{mid}$ and $P_{dil}$ samples
are very similar, showing a maximum at approximately 350 nm, whereas
the absorbance of these samples decreases as wavelength increases.
On the other hand, $C_{mid}$ and $C_{dil}$ samples also show similar
UV-Vis absorption behavior, with a maximum at approximately 500 nm
and with negligible absorbance in the NIR region of the spectra. By
comparing with $P_{mid}$ and $P_{dil}$ it is evident that the main
difference is the shift to longer wavelength of the main band
position. Finally, $P_{conc}$ shows a broad absorption peak with a
maximum at approximately 550 nm and negligible absorbance above 750
nm, whereas the UV-Vis absorption spectrum of $C_{conc}$ is the most
different of all the analyzed samples because the absorbance is
minimum at 450-600 nm and tends to increase in the NIR region.

It is seen in literature that the UV-Vis absorbance of PEDOT can be
used to study doping, oxidation level and conjugation length of this
conducting polymer. In general, three major absorption bands can be
identified in the PEDOT UV-Vis spectrum
\cite{Im2007,Tolstopyatova2009,Lapkowski2000}. The first band
appears in the visible-region (350 - 600 nm) and corresponds to a
$\pi - \pi^*$ electronic transition, associated with the reduced
state of PEDOT \cite{Tolstopyatova2009}, also ascribed to neutral,
dedoped or oligomeric PEDOT segments, and whose intensity diminishes
as the polymer is oxidized \cite{TranVan2001}. It is worth
mentioning that the position of this band shifts to shorter
wavelengths as the conjugation length of the polymer chains
diminishes \cite{Gribkova2016, Gribkova2016b, Garreau2001a}. The
other two absorption bands, which appear in the NIR-region,
correspond to oxidized PEDOT fragments: one, at intermediate
wavelength (800-900 nm), which is attributed to electron transitions
in polarons in the doped state \cite{Gribkova2016}. It appears
during the doping process of PEDOT and, therefore, is associated
with polaron type carriers \cite{Damlin2004}; the other one, at
longer wavelength (1700-1800 nm), which is attributed to bipolaron
states of highly-doped PEDOT, and appears as a “free carrier
tail” related to highly delocalized electrons \cite{Han2006}.
Thus, PEDOT segments with a higher doping level tend to show
absorbance in the NIR-region of the spectrum, whereas more neutral
segments exhibit shorter wavelength absorption bands, specifically
in the visible region of the spectrum.

It is worth noting that, unlike in a PEDOT film where the oxidation
state of the entire polymer sample is controlled by the applied
potential, a chemical synthesis may result in a combination of PEDOT
segments with different oxidation states and conjugation lengths,
which can explain that certain samples do not have a clear narrow
peak or band, instead showing absorbance over a wider range of
wavelengths. Despite this, there are evident changes in the UV-Vis
absorption spectra of the synthesized samples as DBSA and EDOT molar
concentrations increase, and also with the incorporation of
 CoFe$_2$O$_4$  NP. First, it can be inferred that in most of the
samples, with the exception of $C_{conc}$, PEDOT is mainly in its
reduced or non-doped state since the only band present is the wide
one associated with the $\pi - \pi^*$ electronic transition, and
negligible absorbance is observed in the NIR-region of the spectra.
It is an interesting result since several authors reported that it
is necessary to use non-oxidative synthetic routes in order to
chemically obtain PEDOT in its non-doped state \cite{Chiu2005,
Yamamoto1999}. Therefore, our UV-Vis results show that it is
possible to obtain PEDOT in its reduced state via a conventional
chemical route, by using diluted synthetic conditions. Moreover, a
polymer with different doping levels can be prepared by varying the
synthetic conditions, such as EDOT and DBSA molar concentrations,
and with the incorporation of  CoFe$_2$O$_4$  NP in the reaction
medium. Besides, in $C_{dil}$ and $C_{mid}$ samples, the position of
the $\pi - \pi^*$ band is red shifted in comparison to $P_{dil}$ and
$P_{mid}$. This result suggests that, when PEDOT grows in the
presence of   CoFe$_2$O$_4$  NP the obtained polymer is, mainly, in
the same oxidation state, but with polymer chains of longer
conjugation lengths.

It can then be inferred that, for the samples of the present work,
PEDOT in $C_{conc}$ shows the highest oxidation state or doping
level. Then, PEDOT in $P_{conc}$, $C_{mid}$, $C_{dil}$, $P_{mid}$
and $P_{dil}$ samples is essentially in their reduced or non-doped
state and, additionally, the position of the $\pi - \pi^*$
electronic transition in $P_{mid}$ and $P_{dil}$ indicates that, in
those samples, really short conjugation lengths are obtained. This
result was expected because low EDOT and DBSA molar concentrations
do not favor the polymerization of EDOT nor the DBSA doping of the
polymer chains \cite{Choi2004}.

In summary, UV-Vis results indicate that increasing both EDOT and
DBSA molar concentrations in the polymerization medium gives rise to
PEDOT chains with longer conjugation lengths and with better DBSA
doping. The doping role of DBSA, that is, its providing of counter
ions for the polymer backbone, is then evidenced. Besides, the
presence of  CoFe$_2$O$_4$  NP favors the polymerization of EDOT, a
fact that is evidenced by the shorter conjugation lengths obtained
in $P_{dil}$ and $P_{mid}$ in comparison to $C_{dil}$ and $C_{mid}$.
Moreover, the effect of the inclusion of the NP in the
polymerization of EDOT strongly depends on the DBSA molar
concentration employed. On the one hand, in diluted and intermediate
synthetic conditions the main effect is the obtention of PEDOT
chains with longer conjugation lengths, but mainly in the reduced or
neutral/non-doped state as the pristine PEDOT samples. On the other
hand, UV-Vis spectrum of $C_{conc}$ differs significantly from that
of $P_{conc}$, showing an increasing absorbance in the NIR-region of
the spectrum which denotes the presence of polarons. This behavior
indicates that the inclusion of  CoFe$_2$O$_4$  NP also favors the
obtention of oxidized PEDOT fragments, mainly when sufficient DBSA
is added in the polymerization medium. These effects of the
synthesis conditions on the structure of the PEDOT are further
studied in the analysis of FTIR and Raman spectroscopy measurements.

\subsection{FTIR spectroscopy}

\begin{figure}[h]
    \centering
    \includegraphics[width=\textwidth]{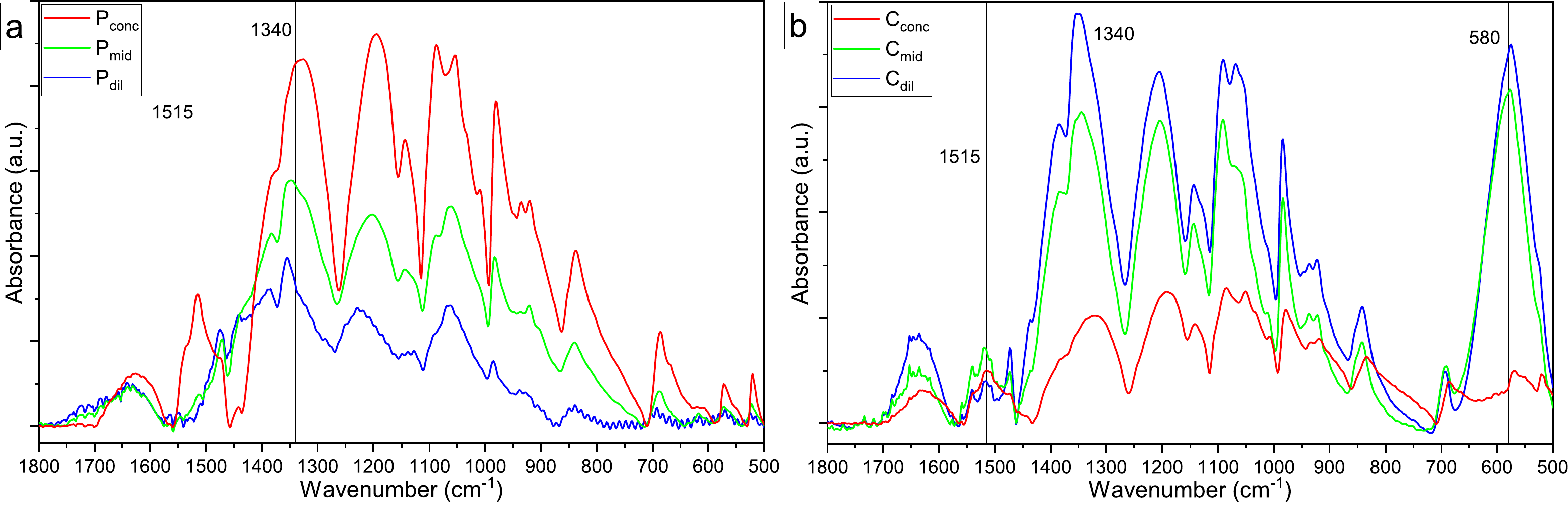}
    \caption{FTIR spectra of: (a) Pure PEDOT samples $P_{conc}$, $P_{mid}$ and $P_{dil}$; and (b) PEDOT: CoFe$_2$O$_4$  composites $C_{conc}$, $C_{mid}$ and $C_{dil}$.}
    \label{fig:FTIR}
\end{figure}

In Fig. \ref{fig:FTIR}, the FTIR spectra of the pure PEDOT and
PEDOT: CoFe$_2$O$_4$  composite samples are shown and, in the
following, the most relevant bands of both PEDOT and
 CoFe$_2$O$_4$  FTIR spectra will be discussed. Firstly, for the
composites, a broad band can be seen near 580 cm$^{-1}$, which is
absent in the pure polymer samples. This band is attributed to
 CoFe$_2$O$_4$  NP, particularly to the Fe-O stretching mode
\cite{Gillot1983,LanusMendezElizalde2020,Antonel2015} and,
therefore, it indicates the presence of  CoFe$_2$O$_4$  NP in the
composites. Besides, its intensity, relative to the other bands of
the spectrum, increases as both EDOT and DBSA molar concentrations
in the reaction medium decrease. This result confirms that with the
diluted synthesis conditions, $C_{dil}$, a composite with a higher
proportion of  CoFe$_2$O$_4$  NP is obtained, which is in agreement
with SEM and TEM observation.

In the range 1800-600 cm$^{-1}$, all FTIR spectra show the typical
bands of PEDOT, indicating the presence of this polymer in all the
synthesized samples. First, the vibrational bands at around 1200,
1150 and 1050 cm$^{-1}$ originate from C-O-C bond stretching in the
ethylenedioxy group, while bands at 980, 840 and 690 are attributed
to vibrational modes of the C-S bond in the thiophene ring. Besides,
the two main bands of PEDOT, located at 1515 and 1340 cm$^{-1}$,
correspond to the C=C and C-C stretching of the thiophene ring,
respectively \cite{Paradee2013, LanusMendezElizalde2020, Yang2007,
HeydariGharahcheshmeh2020, Drewelow2020}. Specifically, the band at
1515 cm$^{-1}$ is ascribed to the presence of the quinoidal
structure of PEDOT \cite{HeydariGharahcheshmeh2020} and, besides,
the increase in its intensity, with respect to that of the 1340
cm$^{-1}$ band, indicates that the polymer chains have longer
conjugation lengths \cite{Drewelow2020, Im2007}. From the FTIR
spectra of Fig. \ref{fig:FTIR} it is evident that the
$A_{1515}/A_{1340}$ absorption ratio is strongly affected by EDOT
and DBSA molar concentrations, and also by the presence of
 CoFe$_2$O$_4$  NP. On one hand, by comparing PEDOT samples (Fig.
\ref{fig:FTIR}a), $P_{conc}$ shows the highest $A_{1515}/A_{1340}$
absorption ratio, whereas in $P_{mid}$ and $P_{dil}$ the band at
1515 cm$^{-1}$ is almost undetectable. The low intensity of this
band has been observed in undoped oligomeric PEDOT
\cite{TranVan2001, Im2007}, thus indicating short conjugation
lengths and non-doped state in $P_{mid}$ and $P_{dil}$, in agreement
with our UV-Vis results. Therefore, FTIR results indicate that
increasing both EDOT and DBSA molar concentrations in the
polymerization of PEDOT gives rise to a polymer with higher
proportion of quinoidal units, together with chains of longer
conjugation lengths. In PEDOT: CoFe$_2$O$_4$  composites (Fig.
\ref{fig:FTIR}b) the same trend in the $A_{1515}/A_{1340}$
absorption ratio is observed as both EDOT and DBSA molar
concentrations increase. More interesting, and unlike $P_{mid}$ and
$P_{dil}$, $C_{mid}$ and $C_{dil}$ show a notorious band at 1515
cm$^{-1}$, which indicate that the presence of  CoFe$_2$O$_4$  NP in
the reaction medium favors the polymerization of EDOT, giving rise
to polymer chains of longer conjugation lengths and higher
proportion of quinoidal units. Moreover, $C_{conc}$ shows the
highest $A_{1515}/A_{1340}$ absorption ratio of all the studied
samples, which could be attributed not only to longer conjugation
lengths and higher proportion of quinoidal units, but also to a
higher doping level of PEDOT \cite{Drewelow2020, Kvarnstrm2000}, as
discussed in the previous UV-Vis section. It is also worth
mentioning that all samples show a peak at approximately 1470
cm$^{-1}$, which is related to benzoid units in PEDOT
(Gharahcheshmeh). As this peak is more intense and better resolved
in the FTIR spectra of the samples prepared by using the diluted and
medium synthesis conditions, it can be concluded that the lower the
DBSA and EDOT molar concentration, the higher the proportion of
benzoideal units in PEDOT. This is in agreement with a low
absorbance at 1515 cm$^{-1}$, that is, with a low proportion of
quinoidal units, as discussed above.

In summary, FTIR spectra show that both EDOT and DBSA molar
concentrations and the presence of  CoFe$_2$O$_4$  NP play an
important role in the polymerization of EDOT, determining the
morphology, the structure and the doping level of the obtained
polymer. In addition to the catalytic effect of  CoFe$_2$O$_4$  NP
on the polymerization of EDOT, it is evidenced that they promote the
obtention of a polymer with a higher doping level, in agreement with
our UV-Vis results.

\subsection{Raman spectroscopy}

\begin{figure}[h]
    \centering
    \includegraphics[width=\textwidth]{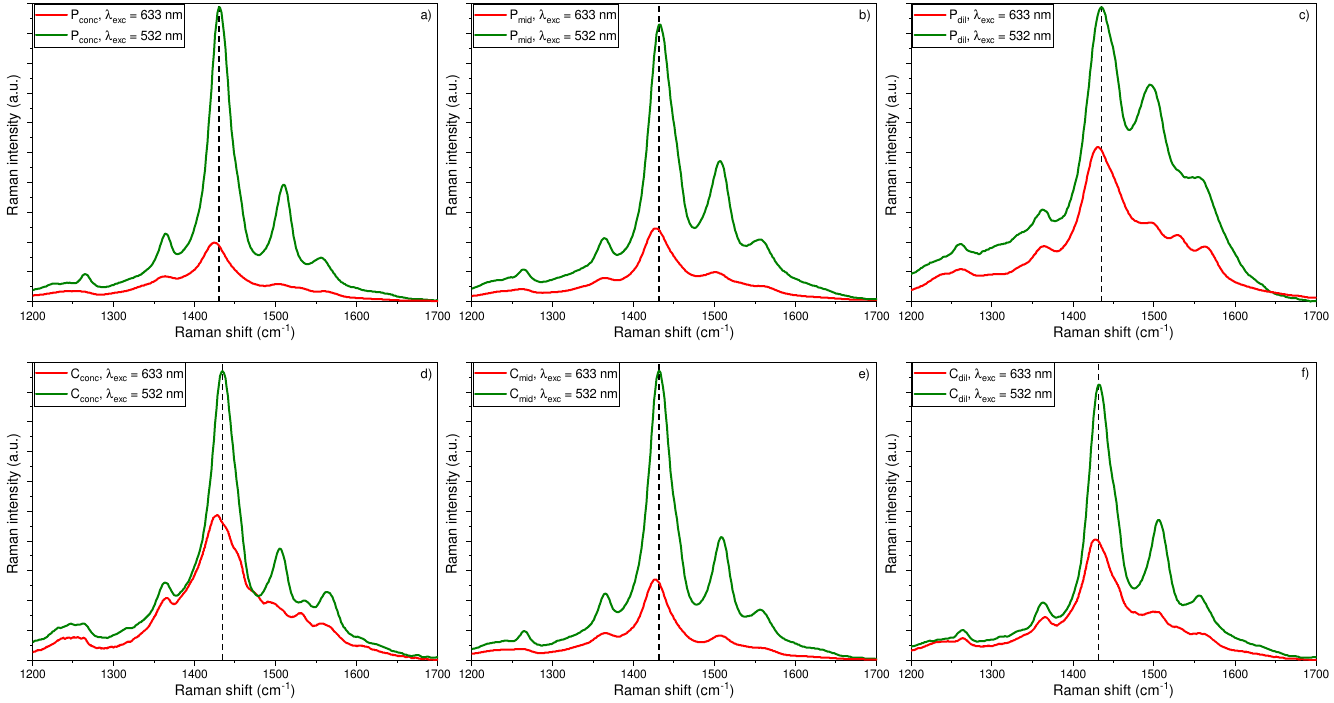}
    \caption{Raman spectra of the pure PEDOT samples (a) $P_{conc}$, (b) $P_{mid}$ and (c) $P_{dil}$; and PEDOT: CoFe$_2$O$_4$  composites (d) $C_{conc}$, (e) $C_{mid}$ and (f) $C_{dil}$, measured at different excitation wavelengths, $\lambda_{exc}$.}
    \label{fig:Raman_laser}
\end{figure}

Fig. \ref{fig:Raman_laser} shows Raman spectra ($\lambda_{exc}$ =
532 and 633 nm) of PEDOT and PEDOT: CoFe$_2$O$_4$  composites, in
the 1200-1700 cm$^{-1}$ range. Raman spectra of all samples are
dominated by the bands in the region of C=C stretching as expected,
particularly by the most intense peak centered at approximately 1425
cm$^{-1}$ assigned to the symmetric C$_\alpha$=C$_\beta$ stretching
vibration on the thiophene rings of PEDOT chains \cite{Garreau1999}.
The position and the width of this peak are particularly sensitive
to the conjugation length, to the molecular environment of polymer
chains and to the oxidation state (redox doping) of PEDOT
\cite{Jorge2021, Salim2024, Tumov2023, Im2007, Garreau1999,
Garreau2001, Chiu2005}.

It is well-known that the position of the Raman bands of PEDOT, in
particular of the most intense band at approximately 1425 cm$^{-1}$,
depends on the laser excitation wavelength, due to the existence of
a $\pi$-bonding system in the polymer \cite{Garreau2001, Kong2022}.
The shift in the position of this band seems to be the result of
several contributions with different signs. Considering firstly the
UV-Vis and FTIR spectra, our samples differ on the conjugation
length and on the oxidation state of PEDOT. Regarding the effect of
the conjugation length on the position of the main band, shorter
wavelengths ($\lambda_{exc}$ = 532 nm in our case) enhance segments
with shorter conjugations lengths, whereas longer wavelengths
($\lambda_{exc}$ = 633 nm in our case) enhance segments with larger
ones \cite{Garreau2001, Duvail2002}. This results in a red shift of
the position of the most intense band when the excitation wavelength
is increased. In other words, the shifts in the position of the most
intense band in Raman spectra when using different excitation
wavelengths indicate that, in PEDOT polymer, there are chains with
different conjugation lengths. Moreover, and considering only the
effect of the conjugation length on the position of the main band,
when the same excitation wavelength is used, comparison between
different samples is also possible \cite{Kong2022}. For example, if
a certain sample presents its main band at a lower wavenumber than
another sample, it means that in the first one the polymer segments
enhanced by the excitation wavelength used, have longer conjugation
lengths \cite{Tumov2023}. On the other hand, and regarding the
oxidation state of PEDOT, it is reported that, at fixed
$\lambda_{exc}$, a blue shift of the position of the main band takes
place when PEDOT is oxidized \cite{Garreau1999, Kong2022,
Farah2012}. In the following, the position of the band at $\sim$
1425 cm$^{-1}$ at both $\lambda_{exc}$ are presented in Table
\ref{tbl:Raman_peak} and will be discussed in terms of these two
contributions.

\begin{table}[h]
    \centering
    \small
    \caption{\ Raman shift (cm$^{-1}$) for the main band of the Raman spectra of the PEDOT: CoFe$_2$O$_4$  composites and pure PEDOT samples. Raman shifts were measured with a resolution of 0.3
    cm$^{-1}$}
    \label{tbl:Raman_peak}
\makebox[1 \textwidth][c]{  \begin{tabular}{lcccc}
        \hline
        \multicolumn{1}{c}{\multirow{2}{*}{Reactants concentrations}} & \multicolumn{2}{c}{PEDOT:NP composite}              & \multicolumn{2}{c}{Pure PEDOT}                      \\ \cline{2-5}
        \multicolumn{1}{c}{}                                          & $\lambda_{exc}$ = 633 nm & $\lambda_{exc}$ = 532 nm & $\lambda_{exc}$ = 633 nm & $\lambda_{exc}$ = 532 nm \\ \hline
        Concentrated                                                  & 1427.6                   & 1434.8                   & 1425.0                   & 1430.1                   \\
        Medium                                                        & 1427.1                   & 1431.7                   & 1427.1                   & 1431.7                   \\
        Diluted                                                       & 1426.8                   & 1431.7                   & 1431.4                   & 1435.2                   \\ \hline
    \end{tabular}}
\end{table}

First, for the three PEDOT samples, the main band, assigned to the
C$_\alpha$=C$_\beta$ symmetrical stretching vibration, undergoes a
blue shift of $\sim$ 4-5 cm$^{-1}$ with the green laser, in
comparison with the red one. This result indicates that, for each
PEDOT sample, there is a distribution of polymeric chains, which
differ on their conjugation lengths. Besides, for each
$\lambda_{exc}$, the main band shifts to lower wavenumbers as the
concentrations of both EDOT and DBSA increase. As none of these
samples exhibits significant absorbance in the polaron state region
of PEDOT (see Fig. \ref{fig:UV}), it indicates that $P_{dil}$,
$P_{mid}$, and $P_{conc}$ do not differ significantly in their
oxidation state. This, in turn, suggests that increasing the
concentration of these reactants in the polymerization medium does
not affect the doping level of PEDOT, thus the observed red shift in
the position of the main band could be attributed to the presence of
polymer chains with longer conjugation lengths. In summary, Raman
spectra of PEDOT indicate that the synthetic conditions, in
particular the EDOT and DBSA concentrations, have a strong influence
on the conjugation length of the obtained polymeric chains.
Moreover, for each PEDOT sample ($P_{conc}$, $P_{mid}$ and
$P_{dil}$) there is a distribution of conjugation lengths, shown by
the shift that undergoes the main band with the different
$\lambda_{exc}$ employed. Finally, by comparing the $P_{conc}$,
$P_{mid}$ and $P_{dil}$ Raman spectra at each $\lambda_{exc}$ used,
a decrease of both EDOT and DBSA concentrations gives rise to
polymeric chains with shorter conjugation lengths.

Raman spectra of PEDOT: CoFe$_2$O$_4$  composites, $C_{conc}$,
$C_{mid}$ and $C_{dil}$, are also shown in Fig.
\ref{fig:Raman_laser}. In the case of $C_{dil}$, and for both
$\lambda_{exc}$, the position of the main band shifts to lower
wavenumber in comparison with $P_{dil}$. Again, since both of these
samples are essentially in the same redox state, the observed red
shift of the position of the main could be attributed to longer
conjugation lengths in $C_{dil}$, as it was deduced in the previous
sections. It is worth remembering that the only difference in the
synthesis of $P_{dil}$ and $C_{dil}$ corresponds to the absence or
presence, respectively, of  CoFe$_2$O$_4$  NP in the reaction
medium. Therefore, this result suggests that the presence of
 CoFe$_2$O$_4$  NP significantly influences the growth of PEDOT
when employing diluted synthetic conditions, giving rise to a
polymer in the neutral state, but with polymeric chains with higher
conjugation lengths. In the case of the synthesis performed with
concentrated conditions, the opposite behavior is observed: for both
$\lambda_{exc}$, the main band in $C_{conc}$ is blue-shifted ($\sim$
3-4 cm$^{-1}$) with respect to $P_{conc}$. In this case, the UV-Vis
spectra of both samples show that PEDOT in $C_{conc}$ is in a higher
oxidation state than in $P_{conc}$ and, as it was said above, this
situation gives rise to a blue shift of the main band in Raman
spectra. Therefore, it can be deduced that in the case of $C_{conc}$
the presence of  CoFe$_2$O$_4$  NP together with the use of higher
molar concentrations of both EDOT and DBSA promote the formation of
PEDOT in a higher oxidation state, with a resulting blue shift of
the main band of the Raman spectra at both $\lambda_{exc}$. Finally,
by comparing Raman spectra of $C_{mid}$ and $P_{mid}$ the main band
appears at exactly the same wavenumber for both $\lambda_{exc}$.
From UV-Vis and FTIR results it was concluded that both samples are
essentially in the neutral or non-doped state, but the presence of
longer conjugation lengths was evidenced in $C_{mid}$. Therefore,
and considering only the effect of the conjugation length and the
doping level of PEDOT on the the position of the main band of the
Raman spectra, a red shift has been expected. Kong et al
\cite{Kong2022} reported that the morphology of PEDOT also plays an
important role in the position of the main band, which undergoes a
blue shift with the reduction of the grain size. As it was discussed
above in this manuscript, the presence of  CoFe$_2$O$_4$  NP
markedly affects the morphology of the resulting PEDOT, particularly
in the medium synthetic conditions. Although in these last two cases
the analysis of the position of the main band is more complicated,
our UV-Vis and FTIR results support the idea that longer conjugation
lengths are obtained when PEDOT grows in the presence of
 CoFe$_2$O$_4$  NP.

In summary, these results suggest that the presence of
 CoFe$_2$O$_4$  NP has a great influence on the growth of PEDOT
chains, on their conjugation length and on the oxidation state
(redox doping) of PEDOT, with a dependence on the DBSA and EDOT
molar concentration employed in the synthesis. In other words, the
presence of these NP governs the polymerization of EDOT, which is in
agreement with the composites morphology, discussed above in this
manuscript and in our previous works: PEDOT mainly grows around
 CoFe$_2$O$_4$  NP, suggesting that they act as catalyst for EDOT
polymerization.

\begin{figure}[h]
    \centering
    \includegraphics[width=\textwidth]{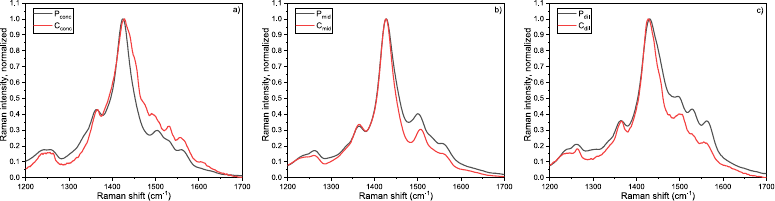}
    \caption{Raman spectra of the pure PEDOT samples and PEDOT: CoFe$_2$O$_4$  composites, measured at excitation wavelength $\lambda_{exc}$ = 633 nm. (a) Concentrated synthesis, (b) medium synthesis, (c) diluted synthesis. Normalized to the maximum of each spectrum.}
    \label{fig:Raman_width}
\end{figure}

In Fig. \ref{fig:Raman_width}, Raman spectra of PEDOT and composites
samples ($\lambda_{exc}$ = 633 nm), normalized to the intensity of
the main peak, are presented. It is reported that the width of the
main band located at approximately 1425 cm$^{-1}$ is related to the
polymer molecular environment \cite{Jorge2021, Lapkowski2000}.
Specifically, the narrowing of the main band indicates that more
bond vibrations take place in a similar molecular environment. In
other words, the narrowing of the broad band suggests that PEDOT
chains are more homogeneously doped. On the other hand, a broadening
of the main band indicates that there are different molecular
environments in the sample, that is, there is no homogeneity in
PEDOT chains doping. With that in mind, changes in the band width
can be rationalized in terms of the DBSA concentration employed in
each synthesis, as follows. $P_{dil}$ and $C_{dil}$ synthesis were
carried out by adding a very low DBSA molar concentration,
specifically 3 times smaller than EDOT molar concentration.
Therefore, a possible interpretation is that at least two types of
PEDOT coexist in the same sample: the first one, with polymer chains
doped by DBSA, and the second one without any doping. When
 CoFe$_2$O$_4$  NP are added to the polymerization medium, EDOT
monomer polymerizes mainly in the vicinity of those NP, a fact that
is evidenced in the SEM image of Fig. \ref{fig:Micros}f. In other
words, PEDOT in $C_{dil}$ is predominantly surrounded by
 CoFe$_2$O$_4$  NP and, therefore, the narrowing of the main band
suggests that in terms of chain doping, this sample is more
homogeneous than $P_{dil}$. On the other hand, Raman spectra of the
samples prepared with concentrated conditions, $P_{conc}$ and
$C_{conc}$, show a broadening of the main band when PEDOT grows in
the presence of  CoFe$_2$O$_4$  NP. This result suggests that the
molecular environment of PEDOT is more homogeneous when
 CoFe$_2$O$_4$  NP are not present in the polymerization medium.
Therefore, in $P_{conc}$, PEDOT is predominantly doped by DBSA,
whereas in $C_{conc}$, two types of PEDOT are obtained, which can be
visualized in the SEM image of Fig. \ref{fig:Micros}d. One type
corresponds to PEDOT doped by DBSA, evidenced in the SEM image as
the region without  CoFe$_2$O$_4$  NP, and the other type
corresponds to PEDOT covering those NP. In the intermediate case of
$P_{mid}$ and $C_{mid}$, the main band undergoes a slight narrowing
due to the presence of  CoFe$_2$O$_4$  nanoparticles. Based on the
contrasts observed in the SEM image in Fig. \ref{fig:Micros}, the
scenario for the diluted concentration seems to persist in a less
pronounced form: in $P_{mid}$, the PEDOT chains remain
inhomogeneously doped, while in $C_{mid}$, PEDOT predominantly grows
around the CoFe$_2$O$_4$  NP, resulting in more homogeneous doping,
though areas with lower doping may still coexist. It is worth
mentioning that the same conclusions can be achieved by comparing
the spectra obtained at $\lambda_{exc}$ = 532 nm (not shown).

\begin{table}[h]
    \centering
    \small
    \caption{\ Raman intensity ratio $I_Q/I_B$ between the quinoid (Q) and benzoid (B) contributions to the main band, for the pure PEDOT samples and PEDOT: CoFe$_2$O$_4$  composites. }
    \label{tbl:Raman_QB}
    \makebox[1 \textwidth][c]{ \begin{tabular}{lcccc}
            \hline
            \multicolumn{1}{c}{\multirow{2}{*}{Reactants concentrations}} & \multicolumn{2}{c}{Pure PEDOT}              & \multicolumn{2}{c}{PEDOT:NP composite}                      \\ \cline{2-5}
            \multicolumn{1}{c}{}                                          & $\lambda_{exc}$ = 633 nm & $\lambda_{exc}$ = 532 nm & $\lambda_{exc}$ = 633 nm & $\lambda_{exc}$ = 532 nm \\ \hline
            Concentrated       & 3.75 $\pm$ 0.51                                 & 14.17 $\pm$ 2.43
            & 12.60 $\pm$ 0.86                              & 10.07 $\pm$ 1.72                   \\
            Medium                      & 3.34 $\pm$ 0.48                                 & 4.30 $\pm$ 0.75
            & 4.58 $\pm$ 1.01                                 & 10.10 $\pm$ 1.94                 \\
            Diluted                        & 2.42 $\pm$ 0.22                             & 2.63 $\pm$ 0.25
            & 2.85 $\pm$ 0.30                               & 4.07 $\pm$ 0.79
    \end{tabular}}
\end{table}

In addition to the information about the conjugation length, the
oxidation state and the homogeneity of the molecular environment of
PEDOT, it is known that the band located at 1425-1435 cm$^{-1}$ has
two contributions due to the different resonant structures of the
conducting polymer \cite{Ouyang2005, Wang2023, Chiu2005}: the
benzoid (B) and the quinoid (Q) structures. It is reported that by
deconvolution and fitting of Raman spectra it is possible to obtain
both contributions to the main band, where the band at lower
wavenumbers is assigned to the Q structures and the band at higher
wavenumbers to the B ones \cite{Ouyang2005, Wang2023, Chiu2005}. The
deconvolution and the fitting for both $\lambda_{exc}$ are shown in
the Supplementary Information (Fig. S3) and, from that, the
intensity ratios $I_Q/I_B$ for each sample were determined, and
presented in Table \ref{tbl:Raman_QB}.

As it can be deduced, the higher the $I_Q/I_B$ ratio, more Q units
are present in PEDOT chains. First, by comparing PEDOT samples for
both $\lambda_{exc}$, the $I_Q/I_B$ ratio increases as both EDOT and
DBSA molar concentrations increase in the reaction medium. This
result suggests that increasing those molar concentrations gives
rise to polymer chains with a higher proportion of Q units, which is
in agreement with other works \cite{Chutia2014}. Another interesting
fact is that, for each sample, the $I_Q/I_B$ ratio is higher when
using the green laser. As it was mentioned above, the green laser
enhances chains with shorter conjugation lengths. Therefore, this
result could indicate that in each polymer sample, PEDOT chains with
shorter conjugation lengths contain more Q units than those with
larger ones. Moreover, considering that in each PEDOT sample the
DBSA molar concentration is fixed, this observation could be related
to a higher doping level of the polymer segments with shorter
conjugation lengths. Surprisingly, the value for $P_{conc}$ at
$\lambda_{exc}$ = 532 nm ($I_Q/I_B$ $\sim$ 14) is much higher than
the estimated at $\lambda_{exc}$ = 633 nm and, also, than those
obtained for $P_{mid}$ and $P_{dil}$ at both $\lambda_{exc}$. This
result suggests that the shortest polymer segments in $P_{conc}$
mainly consist of Q units. As DBSA acts as PEDOT dopant, a higher
DBSA molar concentration results in a higher doping level of the
polymer chains. At this point it is worth remembering that the three
polymer samples are in a similar oxidation state, as it was deduced
before. Therefore, the higher $I_Q/I_B$ ratio for shorter
$\lambda_{exc}$ and as DBSA molar concentration increases could
indicate that a better doping of PEDOT chains gives rise to polymer
segments essentially in the neutral redox state in all cases, but
with a higher proportion of Q units.

By comparing the $I_Q/I_B$ ratios of $C_{dil}$, $C_{mid}$ and
$C_{conc}$ at each $\lambda_{exc}$ a similar behavior is found: the
$I_Q/I_B$ ratio increases as both EDOT and DBSA molar concentrations
increase in the reaction medium. Nevertheless, there are some
differences in comparison with the behavior observed for the polymer
samples. First, the high $I_Q/I_B$ value of $C_{mid}$ at
$\lambda_{exc}$ = 532 nm in comparison with the one observed in
$P_{mid}$ suggests that the polymer segments enhanced with the green
laser in $C_{mid}$ mainly adopt a Q conformation in this case.
Taking into account the UV-Vis spectra of both samples there is not
a substantial change in the oxidation state of PEDOT when
 CoFe$_2$O$_4$  NP are present in the reaction medium. Therefore,
the marked change in the conformation of the polymer chains could be
attributed to the presence of  CoFe$_2$O$_4$  NP and in their effect
on the polymer growth and doping. In fact, SEM and TEM images of
Fig. \ref{fig:Micros}e and \ref{fig:Micros}h showed that, as it was
discussed above, $C_{mid}$ adopts a distinctive morphology with a
major proportion of PEDOT in the vicinity of  CoFe$_2$O$_4$  NP, and
a minor proportion without close NP. This suggests that there are
two types of PEDOT chains in $C_{mid}$: in the one hand, polymer
segments with a shorter conjugation length and with predominantly Q
units, in close contact with  CoFe$_2$O$_4$  NP whereas, on the
other hand, polymer segments without close  CoFe$_2$O$_4$  NP, that
show larger conjugation lengths with a distribution of B and Q
units. In the case of $C_{conc}$, this sample shows a high $I_Q/I_B$
ratio at both $\lambda_{exc}$, indicating that all polymer segments
are predominantly in a Q conformation, while for $P_{conc}$ this is
observed only at $\lambda_{exc}$ = 532 nm. Therefore, the presence
of  CoFe$_2$O$_4$  NP promotes that the larger polymer segments
adopt a Q conformation. From SEM and TEM images of Fig.
\ref{fig:Micros}d and \ref{fig:Micros}g, it is evident that, as
discussed above and similar to $C_{mid}$, in $C_{conc}$ two types of
PEDOT are present. But the main difference of this sample arises
from its UV-Vis absorption spectrum, where the absorbance
drastically decreases at $\sim$ 500 nm (neutral state of PEDOT), and
tends to increase from $\lambda$ = 650 nm to longer wavelengths
(polaron state of PEDOT). Therefore, the incorporation of
 CoFe$_2$O$_4$  NP gives rise to a more oxidized PEDOT,
specifically in a polaron redox state, which is responsible for the
Q conformation of the polymer chains. Finally, for $C_{dil}$, the
$I_Q/I_B$ ratio values obtained at both $\lambda_{exc}$ indicate
that in all polymer segments there is a distribution of B and Q
units. However, a slight increase in the $I_Q/I_B$ ratio at
$\lambda_{exc}$ = 633 nm is observed, suggesting that, also in this
sample, the presence of  CoFe$_2$O$_4$  NP favors the Q conformation
of PEDOT chains.

In summary, these results indicate that the incorporation of
 CoFe$_2$O$_4$  magnetic NP to the reaction medium has a
remarkable influence on the conjugation length, on the oxidation
state and on the proportion of the different resonant structures of
PEDOT. More interesting, the influence of  CoFe$_2$O$_4$  NP on the
proportion of Q units depends on EDOT and DBSA molar concentrations
and, also, on the conjugation length of PEDOT chains, since in each
sample the $I_Q/I_B$ ratio varies with $\lambda_{exc}$. In the cases
of $C_{dil}$ and $C_{mid}$, the main role of  CoFe$_2$O$_4$  NP
seems to be the increase of the conjugation length and the doping of
PEDOT chains, giving rise to polymer segments mainly in a neutral
redox state but with a high proportion of Q units. On the other
hand, in $C_{conc}$ the presence of  CoFe$_2$O$_4$  NP promotes a
PEDOT in the polaron redox state, where polymer segments with
different conjugation lengths adopt mainly a Q conformation.

\subsection{Electrical conductivity}

\begin{table}[h]
    \small
    \caption{\  Electrical conductivity of pure PEDOT samples and PEDOT:NP composites}
    \label{tbl:Cond}
\makebox[1 \textwidth][c]{  \begin{tabular}{lll}
        \hline
        Reactants concentrations    & Pure PEDOT conductivity (S/cm) & PEDOT:NP composite conductivity (S/cm) \\ \hline
        Concentrated  & (1.85 $\pm$ 0.33) 10$^{-2}$            & (2.13 $\pm$ 0.34) 10$^{-2}$                    \\
        Medium        & (2.68 $\pm$ 0.38) 10$^{-5}$             & (2.04 $\pm$ 0.54) 10$^{-3}$                    \\
        Diluted  & \textless 10$^{-6}$                 & (3.47 $\pm$ 0.56) 10$^{-5}$
    \end{tabular}
    }
\end{table}

In Table \ref{tbl:Cond}, the measured electrical conductivity for
the pure PEDOT samples ($P_{conc}$, $P_{mid}$, $P_{dil}$) and the
PEDOT: CoFe$_2$O$_4$  composites ($C_{conc}$, $C_{mid}$, $C_{dil}$)
is shown. Firstly, it is seen that, by comparing the values for
$P_{conc}$, $P_{mid}$, $P_{dil}$ , on the one hand, and those of
$C_{conc}$, $C_{mid}$, $C_{dil}$ on the other, the electrical
conductivity increases when both DBSA and EDOT molar concentration
also increase. Regarding the effect of the DBSA molar concentration
on the electrical conductivity of PEDOT samples, this behavior was
also found by Choi et al \cite{Choi2004} when APS is used as oxidant
in the polymerization of EDOT. It is related to a higher DBSA doping
of the polymer chains, which increases the electrical conductivity.
The increase in the electrical conductivity of PEDOT as DBSA molar
concentration increases was also reported by Chutia et al
\cite{Chutia2013,Chutia2014}. They concluded that the higher the
DBSA concentration is, more DBSA anions can be incorporated between
PEDOT chains, a situation that favors the ordering of those polymer
chains, and which also gives rise to a better $\pi$-electron
delocalization, with a higher electrical conductivity. It is noted
that $P_{mid}$ and $P_{dil}$ have the same EDOT molar concentration,
yet in $P_{mid}$ the DBSA molar concentration used in the synthesis
is 3 times higher than in $P_{dil}$, resulting in a more conductive
sample.

Regarding PEDOT: CoFe$_2$O$_4$  composites, each of them has a
higher electrical conductivity than that of its respective pure
PEDOT sample. This result is, at first, unexpected, since
 CoFe$_2$O$_4$  NP have very low conductivity ($< 10^{-6}$ S/cm),
and thus one would expect that incorporating insulating
nanoparticles into a conductive matrix would result in a composite
less conductive than the pure polymer. It is well known that the
inclusion of insulating or poorly conductive NP in a conductive
matrix (e.g., PEDOT) can hinder charge transport by acting as
obstacles and/or disrupting interchain connectivity, that is, the
percolation pathways. In fact, in our physical mixtures, the
electrical conductivity decreases monotonically as the NP content is
increased, from (8.3 $\pm$ 1.3) $10^{-3}$ S/cm for $M_{13}$ to (4.6
$\pm$ 0.6) $10^{-5}$ S/cm for $M_{90}$, indicating that the NP only
provide an obstacle for the charge transport. Therefore, these
results indicate that the presence of $\mathrm{CoFe_2O_4}$ NP can
limit the polymer ability to form continuous conductive networks
across the material, thus impairing charge transport efficiency.
Thus, these results exhibit the differences between incorporating NP
during the polymerization reaction and mixing them with an already
formed polymer. Specifically, these observations suggest that PEDOT
grows in a different way when  CoFe$_2$O$_4$  magnetic NP are
present in the polymerization medium. Therefore, it is postulated
that, in the present PEDOT: CoFe$_2$O$_4$  composites, the
 CoFe$_2$O$_4$  NP do not merely to provide an obstacle for the
transport of charge carriers, but take an active role in the
polymerization of PEDOT, affecting the physical properties of the
formed polymer.

The increase in the electrical conductivity of PEDOT or other
conductive polymers when they grow in the presence of nanoparticles
was also found in other works \cite{De2009, Gangopadhyay1999,
MuozBonilla2016, Mathad2011, Gu2012}. As it was discussed in the
previous sections, the presence of  CoFe$_2$O$_4$  NP in the
polymerization medium plays a crucial role in the polymer growth,
mainly affecting the morphology of the obtained materials and also
the structure of PEDOT. In particular, the presence of NP leads to
PEDOT materials with a more compact morphology (as evidenced by
SEM), and can promote longer conjugation lengths, higher oxidation
states, and increased quinoid character (as evidenced by UV-Vis and
Raman spectroscopy), all features that typically contribute to
enhanced electrical conductivity. Therefore, the effect of
$\mathrm{CoFe_2O_4}$ NP on the electrical conductivity of the
PEDOT-based composites results from a balance between competing
mechanisms, and the overall electrical conductivity depends on which
mechanism dominates. It is well known that increasing DBSA molar
concentration also increases the polymerization rate and the degree
of polymerization (yield) \cite{Choi2004, Chen2008}. In fact, as
evidenced by SEM, the $C_{conc}$ sample contains the highest
proportion of PEDOT, and some regions clearly exhibit PEDOT domains
that are free of $\mathrm{CoFe_2O_4}$ NP. This indicates that at
least two types of PEDOT coexist in Cconc: one doped by DBSA (in
NP-free regions), and another corresponding to PEDOT covering or
surrounding the NP. Therefore, it is expected that the positive
influence of  CoFe$_2$O$_4$  NP on the physical properties of PEDOT,
particularly on the electrical conductivity, results more prominent
in the composites synthesized with diluted and medium concentration
conditions, since in $C_{conc}$ it is restricted to only a fraction
of the PEDOT. As it was deduced from the analysis of Raman spectra,
for $C_{dil}$, most of the EDOT polymerization occurs near and with
the aid of the nanoparticles, due to the insufficient presence of
DBSA. Conversely, the conditions for $C_{conc}$ allow the formation
of a well doped and adequately grown polymer by itself, and thus the
positive effect of the NP on PEDOT growth, doping and electrical
conductivity is less evident. In other words, it could be
hypothesized that, for conditions that further facilitate the
formation of a more conducting PEDOT, this positive contribution of
the nanoparticles would be negligible. Thus, their most dominant
effect would be to act as physical obstacles that reduce the
electrical conductivity of the polymer \cite{MuozBonilla2016,
LanusMendezElizalde2020, Landa2021}.

It is also reported that the distribution and/or the orientation of
the polymer chains defines the morphology and, therefore, plays also
an important role in the electrical conductivity of conducting
polymer-based samples \cite{Ouyang2005}. From SEM images of Fig.
\ref{fig:Micros}, firstly it was seen that $P_{conc}$ shows a
globular-like morphology with the polymer closely aggregated in
clusters, which allows for better carrier electrical transport
between polymer chains. As the DBSA concentration was reduced, the
morphology became more irregular and flake-like, with more holes and
empty spaces, which hinders conductivity by reducing the free path
of charge carriers through the polymer chains. Therefore, together
with the better DBSA doping and the longer conjugation lengths, the
orientation and the enhanced interaction between polymer chains as
DBSA and EDOT molar concentration increase result in more electrical
conducting PEDOT samples.

The effect of  CoFe$_2$O$_4$  NP on the electrical conductivity of
PEDOT composites can also be partially rationalized in terms of the
observed morphological changes when the polymer grows in the
presence of those NP, with the aid of SEM images. As it was
mentioned above, when  CoFe$_2$O$_4$  NP are present in the
polymerization media, PEDOT grows in a more compacted and ordered
way than in the pure PEDOT samples. Besides, this change in the
composite morphology in comparison to pure PEDOT samples is more
marked in $C_{dil}$ and $C_{mid}$. The observed compactness allows
PEDOT chains to be closer to each other, a situation that improves
their interaction and, therefore, facilitates electrical conduction,
even with the presence of insulating particles among them.

In order to understand, on a molecular level, the conductivity
enhancement as EDOT and DBSA molar concentrations increase, and also
with the incorporation of  CoFe$_2$O$_4$  NP to the polymer network,
spectroscopic results should be considered. It was reported for
PEDOT and for other conducting polymers \cite{Duvail2002} that
longer conjugation lengths improve the transport of charge carriers,
resulting in polymer samples with higher electrical conductivity
values. From Raman analysis of the previous section it was deduced
that increasing both EDOT and DBSA molar concentrations gives rise
to polymeric chains with longer conjugation lengths. It is worth
noting that in both $P_{conc}$ and $P_{mid}$ the EDOT:DBSA molar
ratio is equal to 1, yet in $P_{conc}$ the DBSA and EDOT molar
concentrations used in the synthesis are both 3 times higher than in
$P_{mid}$ resulting in a more conductive sample, presumably due to
the higher conjugation length of its polymer chains. In addition,
UV-Vis results for pure PEDOT samples also suggest the increase in
the conjugation length as both DBSA and EDOT molar concentrations
increase, as it was discussed above in this manuscript. The main
absorption peak attributed to the $\pi-\pi^*$ transition of neutral
PEDOT segments shifts to longer wavelengths, especially in the case
of $P_{conc}$, indicating that this sample is more conjugated than
$P_{dil}$ and $P_{mid}$ \cite{Garreau2001a, Gribkova2016b}. The
incorporation of  CoFe$_2$O$_4$  NP shows a similar effect in the
position of the $\pi-\pi^*$ absorption peak since it appears at
longer $\lambda$ in $C_{dil}$ and $C_{mid}$ in comparison to
$P_{dil}$ and $P_{mid}$, respectively. Therefore, UV-Vis results
also support the idea that the presence of  CoFe$_2$O$_4$  NP in the
polymerization media gives rise to polymer chains with longer
conjugation lengths. In the case of $C_{conc}$ a drastic decrease of
the $\pi-\pi^*$ absorption together with an increase of the
absorbance in the NIR region of the spectra, attributed to electron
transitions in polarons \cite{Gribkova2016b}, is observed. This
result suggests that, in concentrated synthetic conditions, the
presence of  CoFe$_2$O$_4$  NP in the polymerization media promotes
the formation of PEDOT with a higher doping level. In conducting
polymers it is expected that a higher doping level (or redox state)
results in higher electrical conductivity values \cite{Chiu2005}.
Nevertheless, the electrical conductivity values of $P_{conc}$ and
$C_{conc}$ are quite identical, suggesting that the positive effect
of increased oxidation state is counterbalanced by the negative
contribution mainly the disruption of interchain connectivity
consistent with the proposed mechanism. Taking into account that in
$C_{conc}$ the presence of $\mathrm{CoFe_2O_4}$ NP only affects the
PEDOT that grows in close proximity to those NP, we propose that the
structural features of DBSA-doped PEDOT in $C_{conc}$ are similar to
those of PEDOT in Pconc. As a result, the positive influence of the
NP (particularly the higher oxidation states in $C_{conc}$) on the
electrical conductivity is restricted to only a fraction of the
PEDOT in $C_{conc}$. This explains why the negative and positive
effects of the NP effectively counterbalance each other under
concentrated conditions, resulting in the observed similarity in
conductivity between $C_{conc}$ and $P_{conc}$.

Regarding the influence of PEDOT structure on electrical
conductivity, it is also reported that quinoid (Q) structures
facilitate and improve interchain interactions and interchain
transport, resulting in more extended polymer conducting networks
\cite{Ouyang2004, Ouyang2005, Salim2024, HeydariGharahcheshmeh2020}.
This is because the Q structure is associated with a linear or
expanded-coiled conformation, whereas the B structure adopts a coil
conformation. As the linearity increases (Q conformation) the
interaction among PEDOT chains results stronger and, therefore, a
higher proportion of Q units in PEDOT results in higher electrical
conductivity values. From our Raman results it can be concluded that
the increased proportion of quinoid of units (particularly under
diluted and medium DBSA concentrations) correlates with a more
compact morphology and improved structural order, suggesting a
stronger influence of the  $\mathrm{CoFe_2O_4}$ NP on polymer growth
and chain alignment. While the presence of disordered regions cannot
be fully ruled out, the comparative analysis across synthesis
conditions supports the conclusion that the NP contribute to better
structural organization, which in turn enhances electrical
conductivity.

In summary, the electrical conductivity of PEDOT and PEDOT:
CoFe$_2$O$_4$  composites strongly depends on the synthesis
conditions, particularly the EDOT and DBSA molar concentrations and
the presence of  CoFe$_2$O$_4$  nanoparticles in the reaction media.
We propose that both conductivity-enhancing and
conductivity-impairing effects are simultaneously present in all our
composites, with their relative impact modulated by the synthesis
conditions particularly the EDOT and DBSA concentrations. At low
EDOT and DBSA concentrations, the positive effects of the NP on
polymer growth, chain aligment, material compactness, structural
order, conjugation length, doping level, and quinoid character are
more pronounced. This is reflected in the significantly higher
conductivity of $C_{dil}$ compared to $P_{dil}$, and similarly, of
$C_{mid}$ compared to $P_{mid}$. At higher DBSA concentrations
($C_{conc}$), the NP retain their previously observed effects, but
also in this case they promote higher oxidation states in PEDOT.
Nevertheless, these effects are less pronounced than under diluted
and medium synthetic conditions, because they are restricted to only
a fraction of PEDOT, which explains the similar electrical
conductivity values observed for $C_{conc}$ and $P_{conc}$.

\subsection{Electrical transport and conductivity modelling}

In order to deepen our comprehension of how NPs contribute not only
to the polymerization of PEDOT but also to its doping, we will
examine the experimental conductivity results in the polymer and
composites using the general effective medium model (GEM)
\cite{McLachlan1987}. This model determines the effective
conductivity for two conducting media, each characterized by its
electrical conductivity and volume fraction, as indicated in
Equation \ref{eq:gem}:

\begin{equation}
    f_1 \frac{(\sigma_1^{1/t}-\sigma_{eff}^{1/t})}{(\sigma_1^{1/t}+A\sigma_{eff}^{1/t})} + f_2 \frac{(\sigma_2^{1/t}-\sigma_{eff}^{1/t})}{(\sigma_2^{1/t}+A\sigma_{eff}^{1/t})} = 0,
    \label{eq:gem}
\end{equation}
where $\sigma_{1,2}$ and $f_{1,2}$ are the conductivity and the
volume fraction of each medium (1 and 2), respectively.
$\sigma_{eff}$ is the effective conductivity that can be measured,
$t$ is a critical exponent depending on the dimensionality, $f_1 +
f_2 = 1$ and $A = (1 - f_c)/f_c$, where $f_c$ is the percolation
threshold. The typical values considered for a 3D geometry with a
homogeneous distribution of the different phases are $fc = 0.16$, $A
= 5.25$ and $t = 2$. For computational simplicity, we will designate
one media as comprising NP plus the surrounding PEDOT influenced by
potential NP-induced catalysis (PNP), while the other represents the
polymer unaffected by NPs, termed pristine polymer (PP). We will
arbitrarily assign $\sigma_1$ and $f_1$ to the electrical
conductivity and volume fraction of PNP, respectively, and
$\sigma_2$ and $f_2$ to the corresponding parameters for PP. It
should be noted that $\sigma_1$ denotes the effective conductivity
of the PNP regions, which comprise both the insulating NP and the
conducting PEDOT influenced by it. Thus, it should be noted that
this model offers a highly simplified view of the possible real
scenario, in which the division into PNP and PP regions serves as an
effective and functional categorization. These regions may not
correspond to physically discrete domains but instead represent
average areas with distinct conductivities influenced by the
proximity to nanoparticles. Such a conceptual simplification
inherently overlooks possible conductivity gradients near NP
interfaces, as well as particular geometric distributions that may
deviate significantly from a homogeneous arrangement of both phases.

Our approach has been to numerically solve Equation \ref{eq:gem} in
order to determine the dependence of $\sigma_{PNP}$ on $f_{PP}$,
using the experimentally obtained composite conductivities
$\sigma_{eff}$ as fixed parameters. We considered that $\sigma_{PP}$
can range from the value measured in samples containing only PEDOT
to lower values, based on the observed promotion of polymerization
(see SEM observations, Fig. \ref{fig:Micros}) and doping (Raman,
Fig. \ref{fig:Raman_width}) by NP. It should be kept in mind that,
while informative, SEM and Raman measurements do not yield
quantitative details regarding local doping or conductivity
distribution. However, one may reasonably expect that, at minimum,
the modification of $\sigma_{PP}$ in the composite relative to that
of pristine PEDOT would result in a slight decrease when high
reactant concentrations are present. Even so, its doping could be
eventually highly reduced if its polymerization has not been favored
due to being outside the catalyzing action exerted by
 CoFe$_2$O$_4$  NPs on PEDOT. Since the composites were
synthetized nominally with EDOT: CoFe$_2$O$_4$  = 2, and considering
the relative densities (1.33 g/cm$^3$ for EDOT, 5.27 g/cm$^3$ for
 CoFe$_2$O$_4$ ), we can estimate that the nominal volume
fraction occupied by the NP is approximately 20 $\%$. This sets a
maximum value for $f_{PP}$  80 $\%$ that will be considered in the
calculations.

\begin{figure}[h]
    \centering \includegraphics[width=\textwidth]{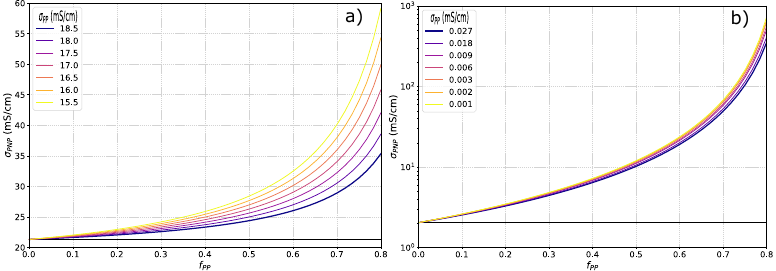}
    \caption{Variation in PNP (PEDOT influenced by  CoFe$_2$O$_4$  NP) phase conductivity with volume fraction of PP (pristine polymer) material in composites corresponding to (a) $C_{conc}$ and (b) $C_{mid}$, for different conductivities of PP. The bolder solid blue curve is obtained by assuming that the conductivity of the PP phase is the same as that of the PEDOT in its synthesis without NPs, while the others correspond to a (a) slightly lower and (b) significantly lower PP conductivity. The horizontal black line indicates the measured experimental conductivity of each composite.}
    \label{fig:Cond}
\end{figure}

In Fig. \ref{fig:Cond}a and \ref{fig:Cond}b, the results obtained
for composites with high and medium reactants concentrations,
respectively, can be observed. It can be noticed that for the
particular case where $f_{PP}$ is zero, and since the entire
composite would be composed of the PNP phase ($f_1 = f_{PNP} = 1 -
f_{PP} = 1$), the conductivity of PNP ($\sigma_{PNP}$) should be
equal to the experimentally obtained value ($\sigma_{eff}$).
Therefore, all curves converge at this point. Additionally, it is
observed that as  $f_{PP}$ increases,  $\sigma_{PNP}$ should also
increase, particularly when $\sigma_{PP}$ adopts the lowest values.
It is interesting to note that for the case of samples with high
reactant concentrations, where we mentioned that, given this excess,
it was reasonable to expect a $\sigma_{PP}$ similar to that obtained
in PEDOT, the variation of $\sigma_{PNP}$ is small across the entire
range of $f_{PP}$ and remains in the order of 35 - 60 mS/cm. In
contrast, for samples with medium reactant concentrations, since
$\sigma_{PP}$ is so low (as derived from the pure PEDOT samples),
$\sigma_{PNP}$ is practically independent of the value of
$\sigma_{PP}$, but it can vary by several orders of magnitude within
the range of variation of  $f_{PP}$.

The values for $f_{PNP}$ and  $f_{PP}$ cannot be  precisely
determined experimentally, since the techniques employed are unable
to distinguish the proportion of pristine PEDOT or PEDOT more
closely associated with NP. Furthermore, it is likely that the
composites are formed by a gradient of increasingly doped and
conductive PEDOT as one approaches the nanoparticles, where
$\sigma_{PNP}$ and $\sigma_{PP}$ would be the average conductivity
for the volume fraction  $f_{PNP}$ and $f_{PP}$, close to the NP and
further from them, respectively. Even though, an approximate value
of these volume fractions can be inferred from the following
considerations: a crude estimation of the lower bound of $f_{PNP}$
corresponds to the volume fraction of the NPs, given that $f_{PNP}$
consists of the NPs and the surrounding PEDOT. Furthermore, SEM
images of the composites can be analyzed, with a threshold
established to differentiate the areas occupied by PEDOT (darker
gray) from those occupied by NP (lighter gray to near white). Since
the diameter of the globular structures in the latter was shown to
be 50 - 70$\%$ larger than that of the NP themselves, it can be
inferred that these lighter areas correspond to particles surrounded
by a PEDOT layer. This volume provides a better estimation of the
lower bound of $f_{PNP}$, which may also be closer to its true
value. Examples of these determinations are shown in Fig. S4.

For the composite $C_{conc}$ , the estimated minimum value of
$f_{PNP}$ is in fact close to the nominal one $\sim$ 20$\%$. Given
that  $f_{PNP}$ is relatively low, we can consider  $\sigma_{PP}$ to
be similar to, or slightly below the conductivity of the pure
polymer (15.5 - 18.5 mS/cm). According to Fig. \ref{fig:Cond}a, this
indicates that $\sigma_{PNP}$ should be between 2 to 4 times
$\sigma_{PP}$. While for the composite $C_{mid}$, the obtained lower
bound of $f_{PNP}$ is (43 $\pm$ 4) $\%$, meaning that $f_{PP}$ is at
most (57 $\pm$ 4) $\%$. Considering Fig. \ref{fig:Cond}b, where
$\sigma_{PP}$ was assumed to be similar to or much lower than that
of the pure polymer (with an arbitrarily selected range of 0.001 -
0.027 mS/cm), $\sigma_{PNP}$ should reach approximately 15 mS/cm.
This value is 8 times higher than the conductivity of composite
$C_{mid}$ and 600 times greater than that of the polymer $P_{mid}$.
This result, achieved within the framework of the GEM model,
highlights the substantial doping effect mediated by the NP.

\section{Conclusions}
Composites based on PEDOT and  CoFe$_2$O$_4$  nanoparticles were
chemically synthesized, and their morphology and polymer structure
was studied and compared to those of pure PEDOT samples in order to
understand their electrical conductivity. First, for PEDOT samples,
it was found that an increase in EDOT and DBSA concentrations in the
synthesis results in higher electrical conductivity values.
Regarding the morphology, concentrated synthetic conditions give
rise to a PEDOT that exhibits an uniform and globular-like
morphology (as opposed to the diluted conditions that results in a
flake-like morphology), which favors the carrier electrical
transport between polymer chains. Regarding PEDOT structure, from
the spectroscopic characterizations, which combined UV-Vis, FTIR and
Raman, it was concluded that higher EDOT and DBSA concentrations
favor the formation of a PEDOT with longer conjugation lengths and
with predominance of quinoid units over benzoid ones, all which also
explain the electrical conductivity enhancement. For the composites,
their unexpected electrical conductivity, which resulted greater
than that of their respective pure PEDOT samples, was as well
explained with the aid of the mentioned characterizations. It was
demonstrated that the nanoparticles play a non-trivial role during
the polymerization of EDOT, and do not merely provide a non
conductive filler, but they actively modify the properties of the
resulting polymer. SEM and TEM microscopies showed that when
 CoFe$_2$O$_4$  NP are present in the polymerization medium,
PEDOT grows in a more compacted and ordered way. This situation
improves the contact and, therefore, the interaction between the
polymer chains, enhancing the electrical conductivity, even with the
presence of insulating particles among them. On the other hand, the
combined spectroscopic techniques showed that the inclusion of NP in
the polymerization medium results in PEDOT with a higher doping
level (or oxidation state), as well as increases its conjugation
length and the proportion of quinoidal resonant structures.
Interestingly, the effect of the NP on the properties of PEDOT
depends on the synthesis conditions, that is, the EDOT and DBSA
concentrations. For example, their conductivity-enhancing effect is
greater for diluted conditions, where the polymer is not adequately
doped by DBSA. This was as well addressed in the structural
characterizations, where it was seen that the role of the NP is
mainly to increase conjugation length and to favour the quinoidal
conformation of PEDOT chains for diluted conditions, while for
concentrated conditions, their promotion of PEDOT to a polaron redox
state is more evident.

The GEM-based conductivity calculations further supported the
depicted conductivity framework for composites, suggesting that to
model their effective conductivity, PEDOT must exhibit two distinct
conductivities: one for the polymer near the  CoFe$_2$O$_4$  NP,
which should be significantly more conductive than the other,
associated with the polymer farther away. For intermediate reactant
concentrations, the polymer adjacent to the NP was estimated to be
up to 600 times more conductive than the PEDOT synthesized without
NP under similar conditions.

In conclusion, this work aims to provide insights into the influence
of  CoFe$_2$O$_4$  nanoparticles on the polymerization of EDOT.
Specifically, it examines how these nanoparticles control the
morphology and structure of the resulting polymer and how these
factors, in turn, affect the electrical conductivity of the
synthesized materials. Additionally, these effects are further
regulated by other synthesis parameters, such as the concentrations
of EDOT and DBSA. The mentioned effects must be considered when
synthesizing composite materials with combining properties of the
precursors, and can be used to tune their physical, chemical,
electrical and magnetic properties, with interesting features beyond
what a simple mixture of components would suggest.

\section*{CRediT authorship contribution statement}
\textbf{Gabriel Paciaroni}: Conceptualization, data curation, formal
analysis, investigation, methodology, software, validation,
visualization, writing original draft, writing – review and
editing. \textbf{Mar{\'{i}}a Ana Castro}: Investigation,
methodology, resources, validation, writing – review and editing.
\textbf{Carlos Acha}: Conceptualization, data curation, formal
analysis, funding acquisition, investigation, methodology, project
administration, software, resources, supervision, validation,
visualization, writing original draft, writing review and editing.
\textbf{Paula Soledad Antonel}: Conceptualization, data curation,
funding acquisition, investigation, methodology, project
administration, resources, supervision, validation, visualization,
writing original draft, writing, review and editing.

\section*{Declaration of competing interest}
The authors declare that they have no known competing financial
interests or personal relationships that could have appeared to
influence the work reported in this paper.

\section*{Data availability}
Data will be made available on request.

\section*{Funding sources}
The authors gratefully acknowledge funding from CONICET (grant PIP
11220200101300CO) and the Universidad de Buenos Aires (grant
20020220200062BA), all of Argentina. P. S. A., C. A. and M. A. C.
are members of the Carrera del Investigador Cient{\'{i}}fico of
CONICET.



 \end{document}